\documentclass[letterpaper]{article}
\usepackage{aaai}
\usepackage{times}
\usepackage{helvet}
\usepackage{courier}
\usepackage{subcaption}
\usepackage[hyphens]{url}
\usepackage{color}
\usepackage{graphicx}
\usepackage{epstopdf}
\usepackage{booktabs}

\def\etal{{\it et al.}}

\begin{document}

\title{Fetishizing Food in Digital Age: \#foodporn Around the World}

\author{}
\author{Yelena Mejova \and Sofiane Abbar \\ Qatar Computing Research Institute, Qatar \\ \{ymejova,sabbar\}@qf.org.qa \And
 Hamed Haddadi \\ Queen Mary University of London, UK \\ hamed.haddadi@qmul.ac.uk }

\maketitle

\begin{abstract}

What food is so good as to be considered pornographic? Worldwide, the popular \emph{\#foodporn} hashtag has been used to share appetizing pictures of peoples' favorite culinary experiences. But social scientists ask whether \#foodporn promotes an unhealthy relationship with food, as pornography would contribute to an unrealistic view of sexuality~\cite{rousseau2014food}. In this study, we examine nearly 10 million Instagram posts by 1.7 million users worldwide. An overwhelming (and uniform across the nations) obsession with chocolate and cake shows the domination of sugary dessert over local cuisines. Yet, we find encouraging traits in the association of emotion and health-related topics with \#foodporn, suggesting food can serve as motivation for a healthy lifestyle. Social approval also favors the healthy posts, with users posting with healthy hashtags having an average of 1,000 more followers than those with unhealthy ones. Finally, we perform a demographic analysis which shows nation-wide trends of behavior, such as a strong relationship ($r=0.51$) between the GDP per capita and the attention to healthiness of their favorite food. Our results expose a new facet of food ``pornography'', revealing potential avenues for utilizing this precarious notion for promoting healthy lifestyles.

\end{abstract}

\section{Introduction}
\label{sec:intro}

\emph{Gastro-porn} was first coined by Alexander Cockburn in 1977 in a review of a cookbook: ``True gastro-porn heightens the excitement and also the sense of the unattainable by proffering colored photographs of various completed recipes''~\cite{cockburn1977gastroporn}. Since then, the rise of diets and fitness in the 80s, complemented by the rise in obesity rates and eating disorders, has sustained the development of food-related media~\cite{oneill2003food}. With such a rich history, it is no surprise the hashtag \#foodporn is one of the most popular hashtags on social media, especially visual media such as Instagram, often accompanied with a close-up of a tantalizing dish. In the age of social media, its users are now defining what food ``pornography'' is to them.



\#Foodporn hashtag is part of a vast lifestyle tracking trend. ``I Ate This" Flickr group, one of the largest and most active on the site, hosts over 640K photos that have been contributed by over 40K members. The practice is so pervasive, manufacturers have released cameras with specific \emph{food} mode\footnote{\url{http://www.cnet.com/news/what-are-all-those-camera-modes-for-anyway/}}, emphasizing the sharpness and saturation of colors. 

Such daily food diaries provide an invaluable resource for food culturalists and public health professionals. In particular, the public sharing of food ``pornography'' contextualizes the favorite foods of social media users around the world. As emotional state~\cite{canetti2002food} and social interaction~\cite{christakis2007spread} have been associated with the healthy weight maintenance, obtaining daily reflections on dietary experience becomes imperative for a holistic view of an individual's relationship with food. The case of \#foodporn is especially interesting, given the possibly negative connotations it introduces, perchance promoting an unhealthy relationship with food, as pornography would contribute to an unrealistic view of sexuality~\cite{rousseau2014food}. In the United States, for example, the most liked Instagram posts are from donut and cupcake shops~\cite{mejova2015foodporn}. In this work, we seek to find whether such attitudes are common across the world.

Instagram is a forum highly suitable for such food log research. We find that a staggering 46\% of the nearly 10 million Instagram posts mentioning \#foodporn we collected were geo-located -- a proportion vastly outnumbering similar collection of Twitter posts (which had only 5.8\% with available geo-location data). The 72 countries we examine prove to have a wide range of integration with international cuisine. Chocolate, cake, and other heavy foods dominate the \#foodporn conversation, with some nations sporting their own unhealthy trends with \#gordice (in Brazil) and \#gourmandise (in France), meaning ``fat'' and ``gluttony'', respectively. However, we hesitate equating the \#foodporn trend to the effects of non-food pornography. Among the most popular foods is salad, followed by potentially healthy alternatives of sushi and chicken. Furthermore, the social approval (in terms of likes and comments), as well as user following is the highest for hashtags associated with the healthy lifestyle. Thus, the impact of the study presented below is two-fold: an introduction of a fertile dataset for health research, and a case study of a concept's re-definition, as it is owned and explored by the global social media community.  

\section{Related Works}
\label{sec:related}

Use of social media for monitoring public health has been increasingly popular in the last few years (for a systematic review, see~\cite{capurro2014use}). Popular Online Social Networks (OSNs) such as Twitter, Instagram, and Facebook, with over a billion users, have become a rich source for social scientists, psychologists, and health professionals. A range of behavioral health issues have been studied using OSNs, including depression and mental health~\cite{park2015manifestation}, obesity and diabetes~\cite{abbar2014you}, tobacco use~\cite{prier2011identifying}, and insomnia~\cite{paul2011you}. 
Complementary to physical activity data and Electronic Health Records (EHRs), social media data provides a record of daily social engagement necessary for a holistic view on individuals' health~\cite{haddadi2015360}. Completing the information cycle, social media has also been illustrated to be a valuable tool for wellness and health promotion~\cite{neiger2012evaluating}.
 
Despite being in use for over 5 years, Instagram has received little attention from the research community. With a rich source of data around the picture posts, including social network of the users, a folksonomy of hashtags~\cite{yamasaki2015revealing}, and an association with a hierarchy of locations, it provides a detailed view of individuals' daily lives and habitual activities. Hu~\etal~\cite{silva2013comparison} present a first study of users' posting behavior on Instagram, showing that food pictures are a major feature of Instagram posts. In recent work, we associate health-related activity captured on Instagram to regional obesity, diabetes and other census statistics in the US~\cite{mejova2015foodporn}, finding distinctions between the behavior of users who are more, or less, likely to be healthy. 

Dietary behavior research calls for a more global view, since much of diet-related behavior is cultural~\cite{counihan2013food}. As Fischler~\cite{fischler1988food} puts it, by selecting and cooking food, one ``transfers nutritional raw materials from the state of Nature to the state of Culture''. Culture-specific ingredient connections have been discovered by Ahn \etal~\cite{ahn2011flavor} who mine recipes to create a ``flavor network''. Temporal nature of food consumption has been explored by West \etal~\cite{west2013cookies}, who mine logs of recipe-related queries. They illustrate the yearly and weekly periodicity in food density of the accessed recipes, with different trends in Southern and Northern hemispheres, suggesting a link between food selection and climate. In this research, we quantify the context around the favorite foods of nations around the world in terms of perceived healthiness or unhealthiness, social situations, and sentiment. We further correlate these behaviors with nation-wide demographics.

The wider use of tantalizing food imagery has been studied by social scientists, often drawing parallels between ``food porn'' and the non-food pornography. Anne McBride recently asked academics and chefs about the term, finding that many associate it with unrealistic, ``sexy'' photographs of food, mostly used for advertisement~\cite{mcbride2010food}. The word itself is meant to attract attention, but may have other connotations, such as it being ``indecent [...] when there is so much hunger in the world''. Our research validates the opinions of these experts, in finding the attitudes every-day social media users associate with this term.



\section{Data}
\label{sec:data}

Our main source of data is Instagram, a social platform for sharing photo and video content via mobile devices (by design containing more geo-centric data compared to desktop-oriented social sites). 
Using the \texttt{Tags} endpoint of Instagram API, we collected all posts containing the keyword \#foodporn, resulting in a collection spanning 6 Nov 2014 -- 6 Apr 2015 
containing 9,378,193 
posts from all over the world. 

\begin{figure}[t]
 \centering
 \includegraphics[width=0.9\linewidth,bb=10 10 430 300]{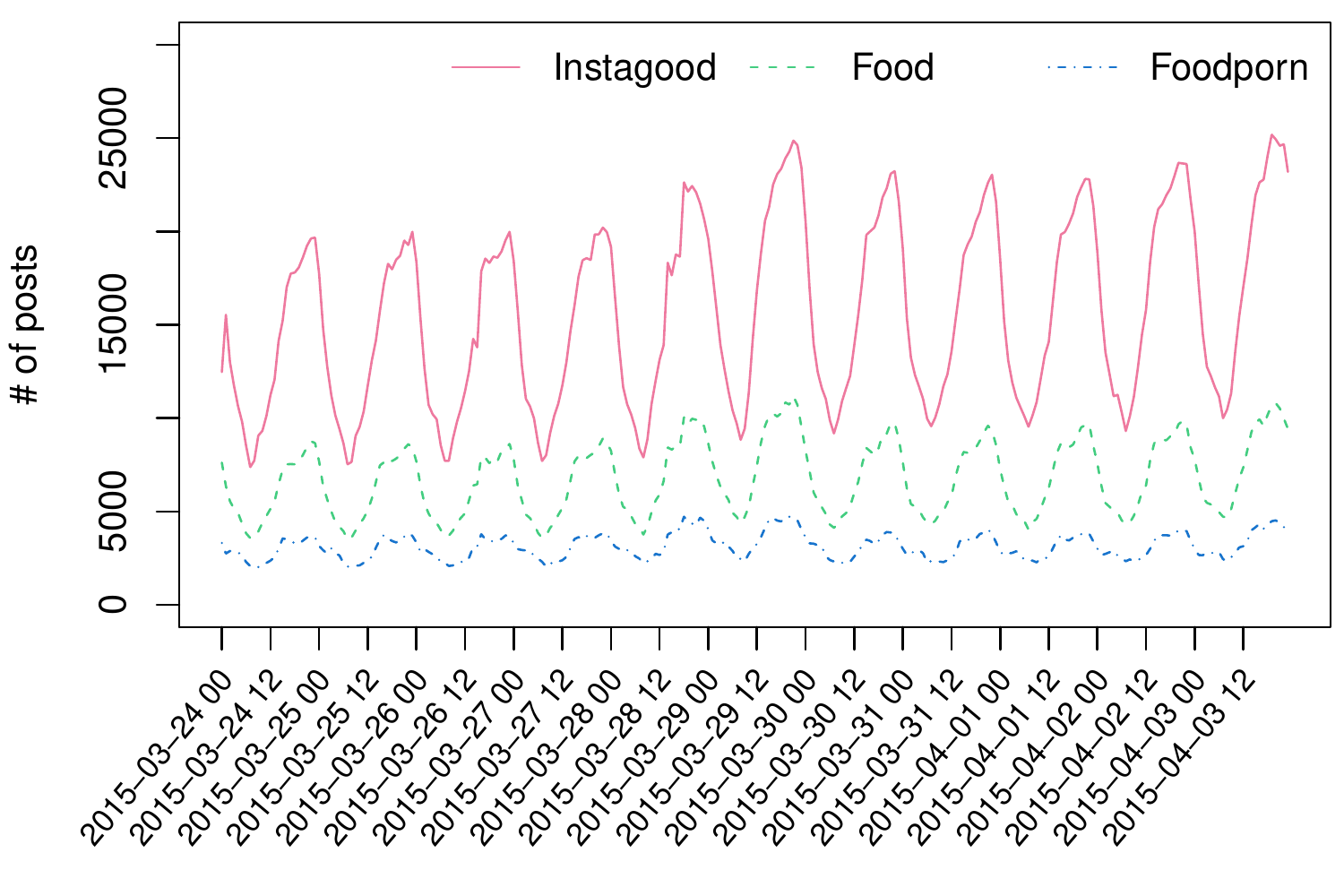}
 \caption{Volume of Instagram posts mentioning \#instagood, \#food, and \#foodporn hashtags.}
 \label{fig:volume_comparison}
\end{figure}

To put it in context of other popular streams, we collected two additional datasets for a smaller overlapping time interval, one to compare to a general food-related conversation, with hashtag \#food (consisting of 1,460,226 posts), and similarly to general discussion on Instagram not necessarily about food, with hashtag \#instagood (3,368,416 posts). Figure \ref{fig:volume_comparison} shows the volume of the three streams, which decrease with more topical specificity.

\begin{table*}[t]
\begin{center}
\caption{Dataset statistics for Instagram datasets for the overlapping period (25 Mar 2015 -- 03 Apr 2015).}
\label{tbl:datasets}
\begin{tabular}{l|rrrrr}\toprule
Hashtag & posts & users & avg posts/user  & \% geotagged posts & unique locations \\ \midrule
\#instagood & 3,368,416 & 920,495 & 3.65 & 28.58\% & 915,122 \\
\#food & 1,460,226 & 619,340 & 2.35 & 31.56\% & 696,640\\
\#foodporn & 675,145 & 322,939 & 2.09 & 42.77\% & 361,653\\ \bottomrule
\end{tabular}
\end{center}
\end{table*}

The posts both in large \#foodporn dataset and smaller ones were then geolocated using the \textit{World Borders shape file}\footnote{http://thematicmapping.org/downloads/world\_borders.php} by converting the (latitude, longitude) pair to the country code from which the post is published. It is worth notice that the geolocation module returns ``None'' when the post does not contain geo-coordinates or when they fall close to some disputed/unclear borders.     
The results of the geo-location can be seen in Table \ref{tbl:datasets}. 
The proportion of successfully located posts increases from \#instagood at 28.6\%, to \#food at 31.6\%, and to \#foodporn at 42.8\%. 
We observe that not only do users geo-tag food-related posts more than generic ones, but especially so the food they particularly enjoy.

\begin{figure}[t]
 \centering
 \includegraphics[width=0.9\linewidth,bb=10 10 440 300]{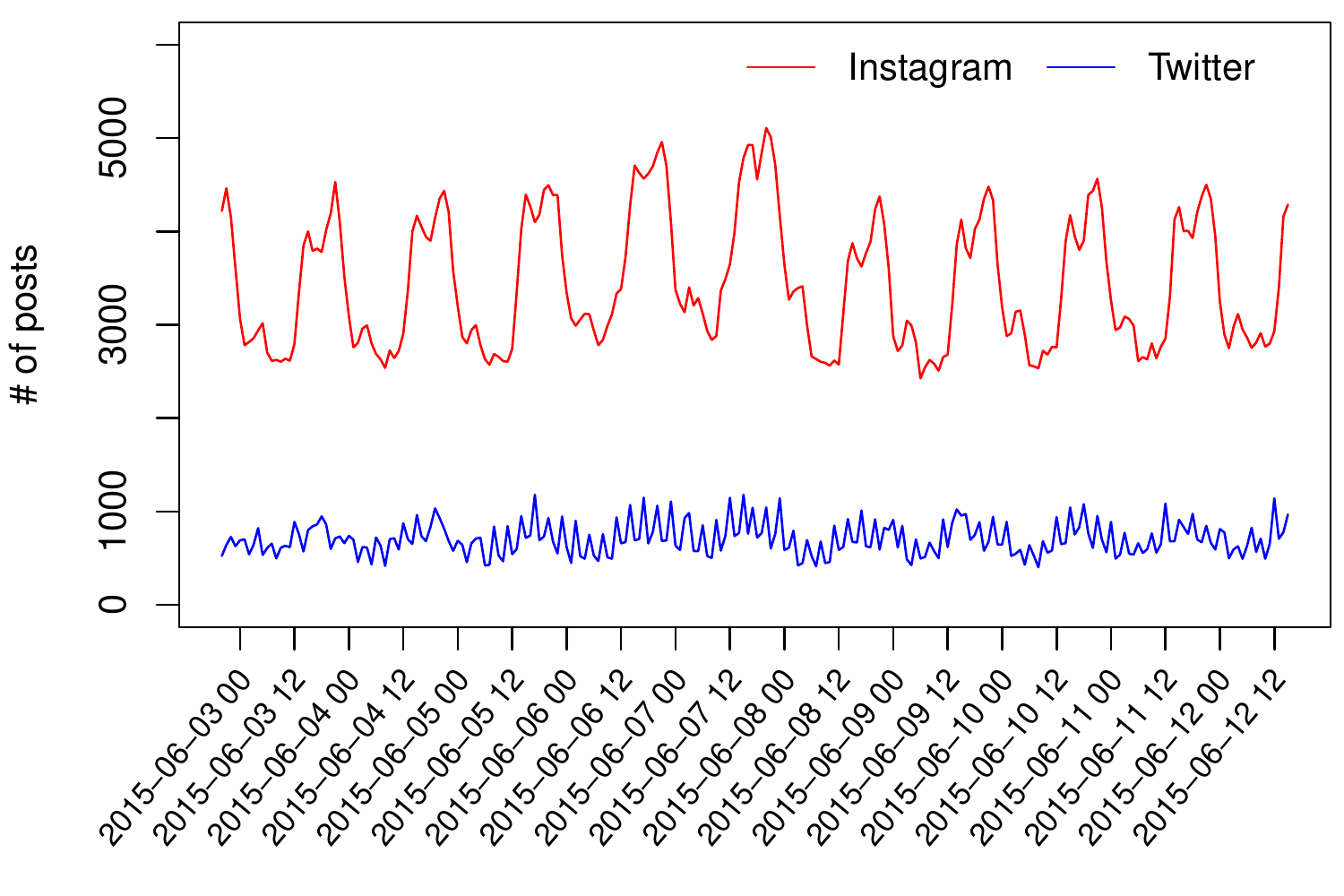}
 \caption{Volume of Instagram and Twitter posts mentioning \#foodporn.}
 \label{fig:instagram_vs_twitter}
\end{figure}

As Twitter is another popular micro-posting platform, we have collected a similar dataset of \#foodporn mentions spanning 10 days 02 Jun 2015 -- 12 Jun 2015. 
Similarely to Instagram posts, we used our geolocation module to map tweets into countries (see Figure \ref{fig:instagram_vs_twitter} for volume statistics).
However, Twitter provided much fewer geo-located posts, with only 5.76\%, compared to Instagram's 46.2\% in the same time span.

The above statistics show \#foodporn Instagram collection to be unique in both the volume (compared to Twitter) and the proportion of geo-located data (compared to, say, \#instagood). Thus, in this study we use the complete \#foodporn Instagram dataset encompassing 5 months, and containing 9.3 million posts.

This data spans 222 countries, with a characteristic long tail distribution. 
Out of these, we select 72 countries having at least 500 unique users in our dataset for further examination. 
At their head are the United States, Italy, and United Kingdom, with the tail including Malta, Iran, and Pakistan.

\section{Prominence and Use}
\label{sec:use}

\begin{figure}[t]
\centering
\includegraphics[width=0.9\linewidth,bb=5 5 590 520]{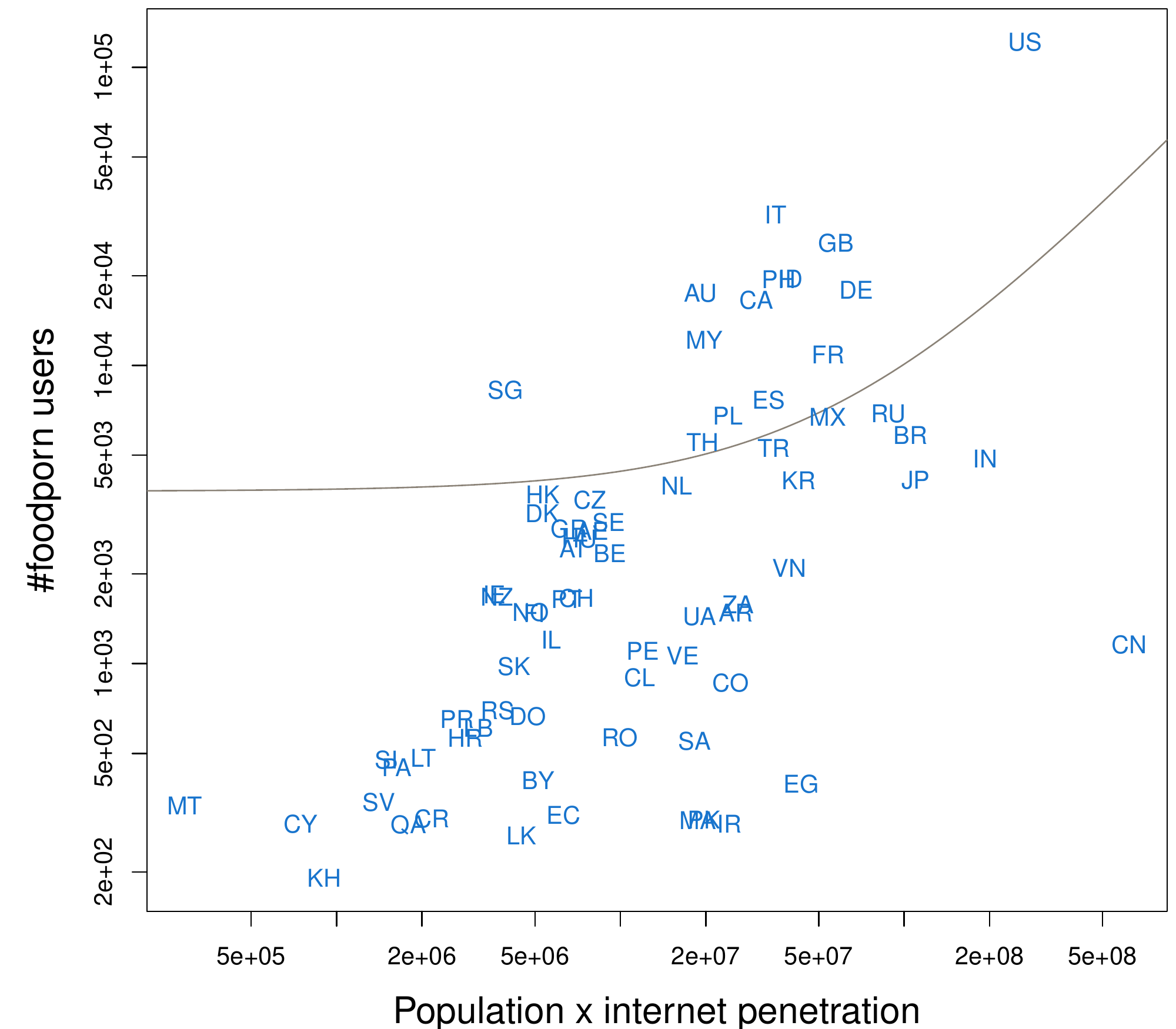} 
\caption{Country population versus the unique number of users posting at least once with \#foodporn hashtag.}
\label{fig:popusers}
\end{figure}

The reach of \#foodporn hashtag across the globe is difficult to understate. In the 150 days of our dataset, over $1.7$ million individual users tagged their posts with it, making the average rate at 62K posts per day. As with most social media, United States dominates the conversation, with Italy being the second-highest user. Figure~\ref{fig:popusers} plots the internet penetration-adjusted country population statistics\footnote{Statistics for 2013 using World Development Indicators data, as described in Demographic Correlation Section.} to the unique number of users posting at least once with \#foodporn from that country, along with a linear regression line (curved, due to log/log axes). We find China (CN) to be an outlier, with disproportionately few users in our dataset compared to its population. Although the top right corner is dominated by European countries, notable exception are Australia (AU), Malaysia (MY), and Singapore (SG) -- the latter especially unusual considering its smaller population. As we show later, Asian countries have by far the largest proportion of users dedicated to tracking their \#foodporn experiences.

\begin{figure*}[t]
\centering
\begin{subfigure}{.3\textwidth}
  \centering
  \includegraphics[width=\textwidth,bb=5 5 790 520]{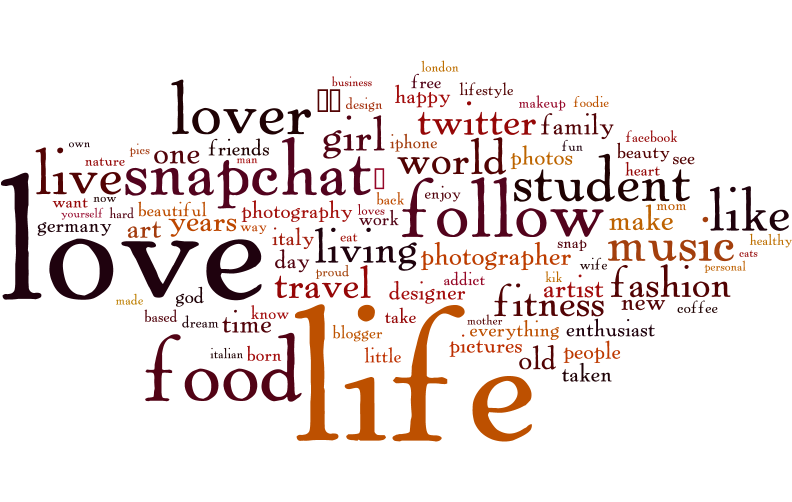}
  \caption{Singletons}
  \label{fig:singletons}
\end{subfigure}\hfill
\begin{subfigure}{.3\textwidth}
  \centering
  \includegraphics[width=\textwidth,bb=5 5 790 520]{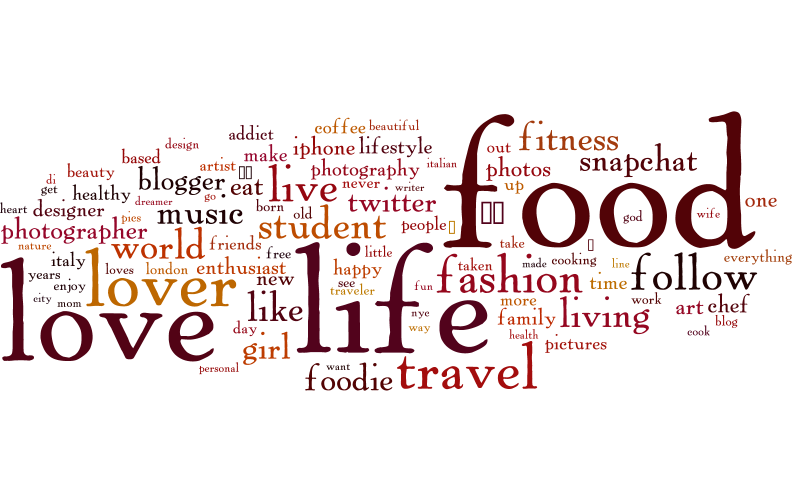}
  \caption{Residents}
  \label{fig:residents}
\end{subfigure}\hfill
\begin{subfigure}{.3\textwidth}
  \centering
  \includegraphics[width=\textwidth,bb=5 5 790 520]{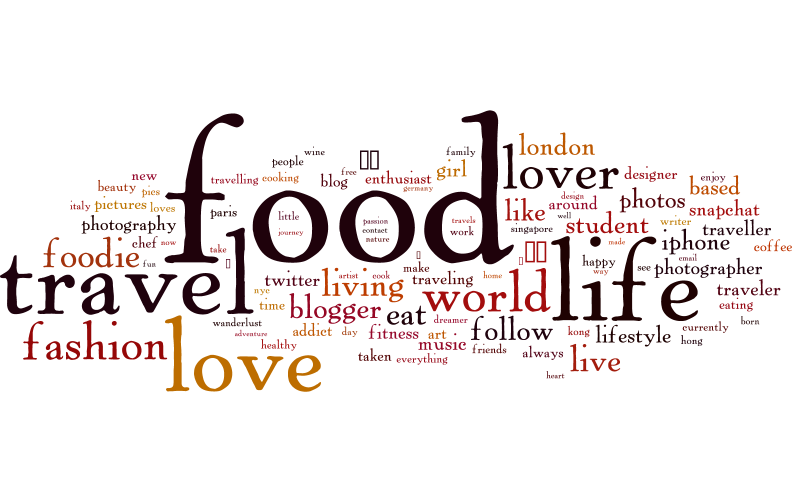}
  \caption{Travelers}
  \label{fig:travelers}
\end{subfigure}
\caption{Biographies of user groups: \emph{singletons} (posting once), \emph{residents} (posting several times from same country), and \emph{travelers} (posting from several countries).}
\label{fig:users_mobility}
\end{figure*}

Next, we focus on the users. We query the \texttt{Users} endpoint of Instagram API to request the profiles of $436,472$ 
(out of $1.7M$ total users) having at least one geo-tagged post with \#foodporn hashtag. Each user profile comes with basic information such as \textit{id}, \textit{username}, \textit{full name}, \textit{bio}, and the numbers of total shared \textit{media}, and their \textit{followers} and \textit{friends}.
We compute then for each user the set of countries associated with all her posts on \#foodporn. 
To differentiate the users by their contribution to the dataset, we define three categories: \textit{Singletons, Residents,} and \textit{Traveleres}. 
The first category contains users who have posted only once in the time frame of our study (\textit{Singletons}). 
The second encompasses users who posted more than once, but whose posts are associated with only one country (\textit{Residents}). 
The third contains users posting from several countries (\textit{Travelers}). 
Note that we cannot tell where a user resides permanently, thus the above definitions are shortcuts to characterize behavior as captured through the use of \#foodporn hashtag. A more thorough examination of user's activity and content is left for future work.

\begin{table}[h]
\begin{center}
\small
\caption{Summary statistics of user groups.}
\label{tbl:users_mobility}
\begin{tabular}{l|rrrrr}\toprule
Category & users & \% users & posts  & \% posts & p/u \\ \midrule
Singletons & 122,581 & 28.09\% & 122,581 & 2.76\% & 1 \\
Residents & 290,741 & 66.61\% & 3,625,784 & 82.24\% & 12.47  \\ 
Travelers & 23,150 & 5.30\% & 687,390 & 15\% & 29.69 \\ \bottomrule
All & 436,472 & 100\% & 4,435,755 & 100\% & - \\ \bottomrule
\end{tabular}
\end{center}
\end{table}

Table \ref{tbl:users_mobility} provides summary statistics of the three user groups. While Travelers constitute only $5.30\%$ of the entire population, they contribute $15\%$ of the posts. In fact, a Traveler user posts in average $2.3$ times more than a Resident user. Note that Singletons may also ``reside'' in the same country as a Resident user, but they are not heavy users of \#foodporn hashtag.

Among the countries we consider, we find Asia to have the greatest dominance of Residents over Singletons, that is, its users are determined to use \#foodporn habitually. These countries include Hong Kong, Korea, Singapore, Taiwan, Thailand, Japan, etc, and their populations have under 20\% singletons. The countries on the opposite side of the spectrum are Norway and Finland, who have over 40\% singletons. In terms of proportion of travelers, Cambodia and China stand out at 27 and 20\%, respectively, suggesting their native populations are not as involved in the conversation.

Next, we examine the profiles of users, particularly their self-identified biographies. 
Figure \ref{fig:users_mobility} shows the most frequent words mentioned by users of each category (with stop words removed). 
Interestingly, we find that Singleton users' interest is about general topics such as \textit{life} and \textit{love} -- these users only occasionally use \#foodporn among their many other interests. 
Residents and Travelers share the same top interest of \textit{food}, while differing in their top second interest which is \textit{travel} for Travelers and \textit{life} for Residents. 

\begin{figure}[t]
\centering
\includegraphics[width=0.9\linewidth,bb=5 5 590 520]{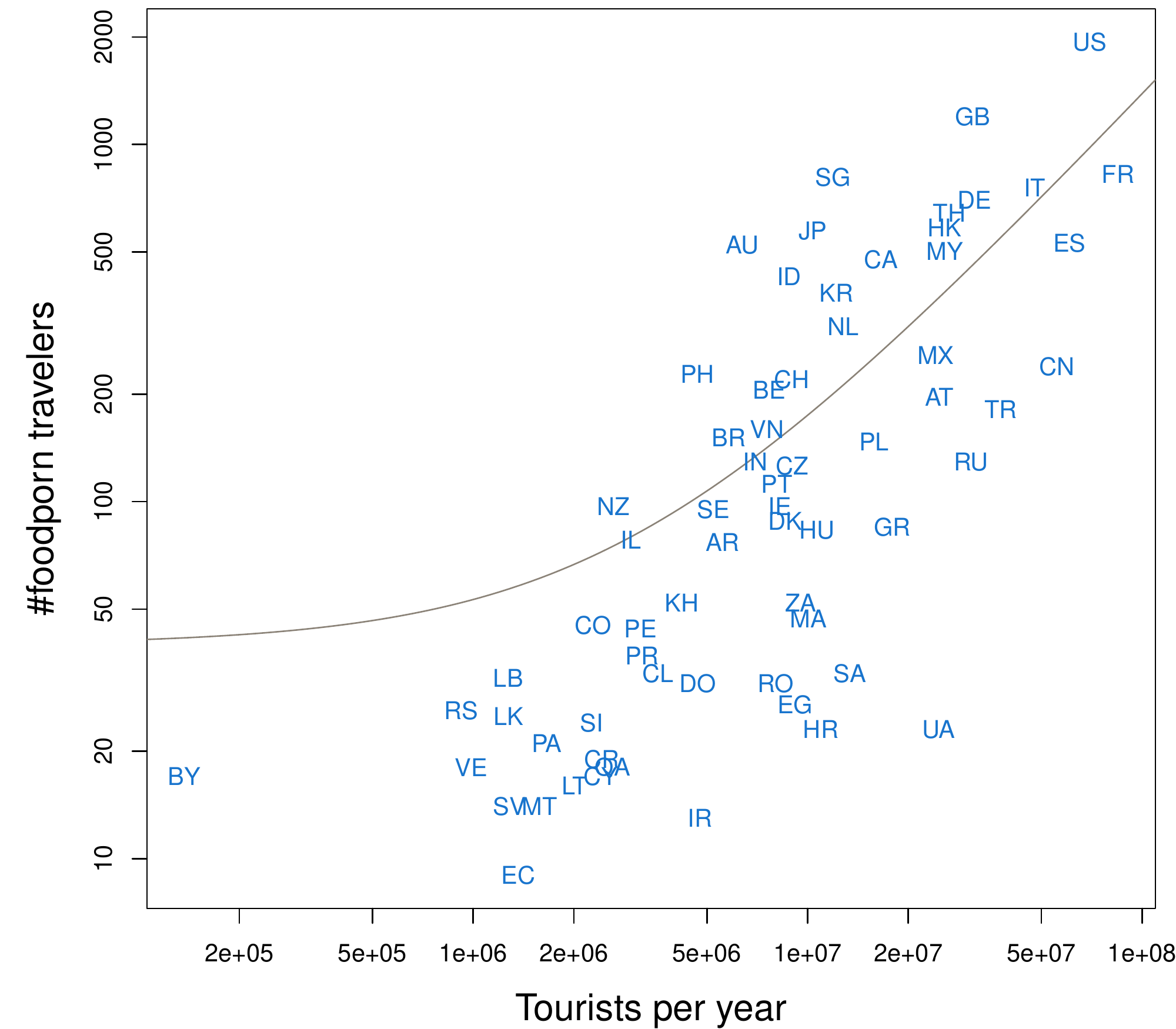} 
\caption{Number of tourists entering country per year versus number of Travelers users posting from a country.}
\label{fig:touristtravel}
\end{figure}

We further find support to the fact that tourism may drive the experiences users associate with \#foodporn when we relate the number of Travelers in our dataset to the number of tourists entering the country. Figure~\ref{fig:touristtravel} shows the relationship, which has strong positive correlation.

As we move forward with the content analysis, in most cases we do not distinguish between the groups above. First, it is impracticable (and often impossible) to locate the ``home'' country of 436K users. Second, as we show in Demographic Correlation Section, countries with expat cultures provide their own characterization of local \#foodporn cuisine. Finally, due to sparsity, it is often impossible to study all three cohorts for all countries. However, the interaction between local and outside perspective on a country's cuisine is an exciting future research direction.

\section{Food Preferences}
\label{sec:foodpref}

Instagram is a media (photo \& video) sharing platform, but it is the hashtags which make it searchable, and which provide valuable annotation of the post's content and context. Our analysis combines qualitative and quantitative insights using this textual information.

\subsection{Global View}

First, we ask, which foods do people in various countries enjoy so much as to associate them with the \#foodporn hashtag? 
We begin by computing the frequencies of all hashtags other than \#foodporn. To prevent prolific users from dominating the rankings, we count the use of each hashtag only once per user. Similarly, we aggregate hashtags per country, first normalizing the frequencies (to probabilities), such that over-represented countries (most notably United States) does not dominate the others. Thus, we find at the top of such a list the tags \#food, \#instafood, \#yummy, \#delicious, and \#foodie. Out of the meals, \#dinner precedes \#lunch, suggesting \#foodporn is more often experienced during the evening meal. 

In particular, we are interested in the foods mentioned in these posts. To identify such foods, we use Crowdflower\footnote{\url{http://crowdflower.com/}} to label the top 100 hashtags of each country as either food or not. To assist labelers with foreign foods, links to Google Translate\footnote{\url{https://translate.google.com/}} and Google Search\footnote{\url{https://www.google.com/}} were provided. Aggregating over 3 labels for each term, 3,870 terms were labeled, out of which 972 mentioned food (with high inter-rater agreement of 91\% label overlap). 

\begin{figure}[t]
\centering
\includegraphics[width=0.8\linewidth,bb=5 5 390 320,trim={2cm 2.2cm 1cm 2.5cm},clip]{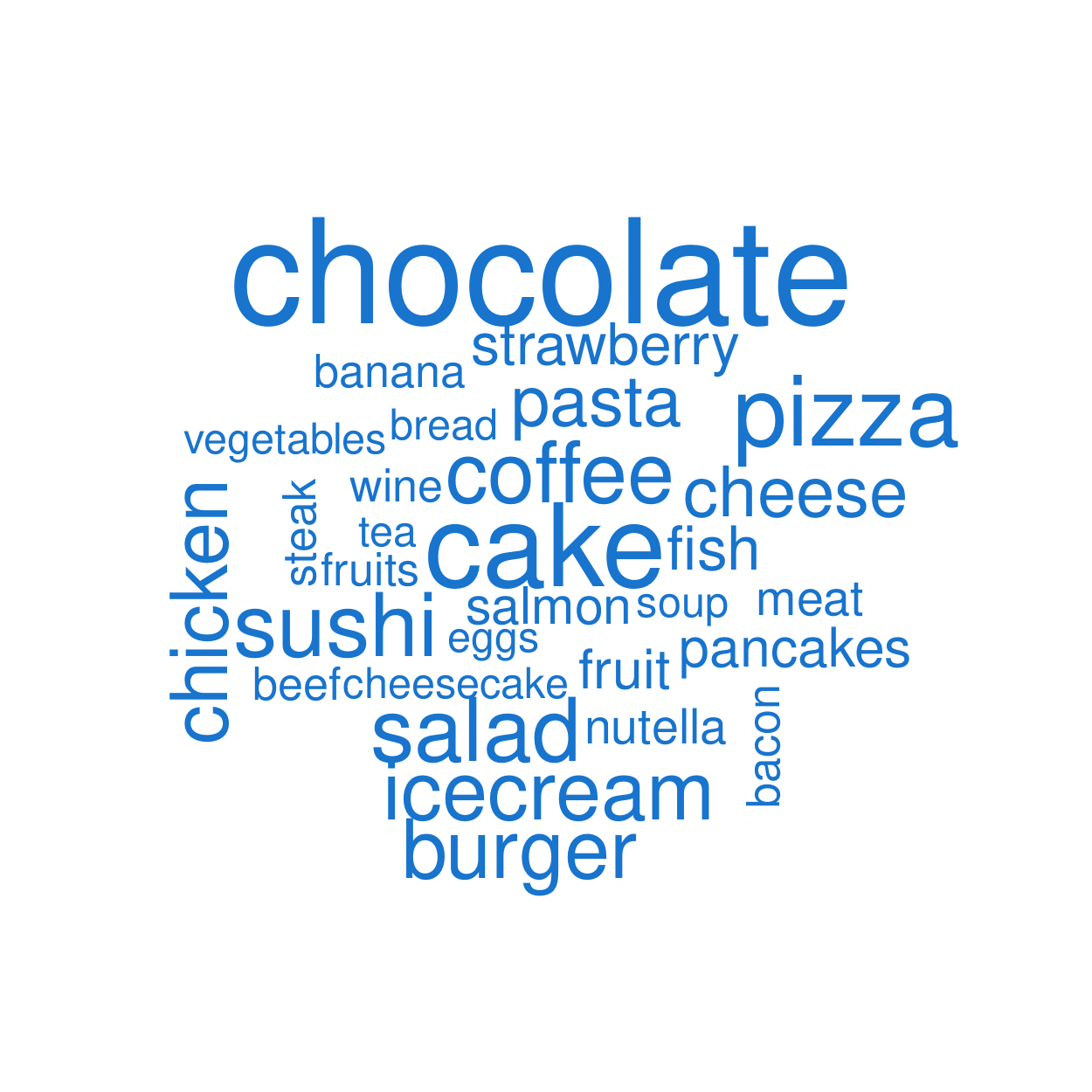} 
\caption{Top foods mentioned in the dataset, normalized by user and country.}
\label{fig:topfoods}
\end{figure}

To compile a global ranking of foods, not dominated by any one country, we normalize hashtag frequencies per country and then aggregate globally. The resulting top 30 foods are shown in Figure \ref{fig:topfoods}. Globally, we find that users are most excited about sweets and fast food. In the 72 countries we consider, when we look at the top 3 foods, \emph{chocolate} appears in 56 (78\%), and \emph{cake} in 29 (41\%). The top non-sweet foods are \emph{pizza}, \emph{salad}, \emph{sushi}, and \emph{burger} -- representing both Western and Eastern cuisines. Top drink is \emph{coffee}, top alcoholic one is \emph{wine}, and top fruit is \emph{strawberry}. Although the vast majority are generic ingredients and dishes, a brand appears at the 18th spot -- \emph{Nutella}, a hazelnut chocolate paste -- which has sold 365,000 tonnes worldwide in 2013\footnote{\url{http://www.bbc.com/news/magazine-27438001}}.

As one can see from Figure \ref{fig:topfoods}, the terms most used alongside \#foodporn are in English (even when volume is normalize per-country such that, for example, United States or Great Britain, do not dominate the ranking). This could be due to some fraction of our users being English-speaking tourists (recall that ``travelers'' contribute 15\% of the posts). However, the fact that English remains the predominant language throughout the nations in our dataset, once again points to it being the \emph{lingua franca} of social media (as discussed, for instance, in \cite{tagg2012role}). Incidentally, we manually labeled the top 50 tags used by the three groups in each country for language, and found a surprisingly consistent proportion of English (at 85\% for Singletons and 88\% for the other two groups). This further emphasizes the prominence in the use of English in both heavy and light social media users.

In order to find foods most used in each country, we compute a score by subtracting the probability a food is mentioned in our dataset from the probability it is mentioned in a particular country. Due to space, we direct the reader to \footnote{\texttt{\url{http://cdb.io/1jSHJTr}}} to explore the top foods of the countries. A selection is listed in Table \ref{tbl:topcountryfoods}. At the top we often find national specialties -- Canadian \emph{poutine}, Ecuadorian \emph{ceviche}, Japanese \emph{ramen}, Swiss \emph{fondue}. Major diasporas can be found, such as the Asian one in Canada (the largest minority at 15\% of population\footnote{\url{http://www12.statcan.gc.ca/nhs-enm/2011/dp-pd/dt-td/Index-eng.cfm}}). Yet other countries show the international nature of their residents, such as Qatar and USA, which have a wide array of cuisines. Finally, we find local alphabets and words, although even those are often transliterated into latin script (as in the case of Iran). 

\begin{table}[t]
\begin{center}
\caption{Top most distinct five foods co-occurring with \#foodporn for select countries.}
\label{tbl:topcountryfoods}
\begin{tabular}{r}
\toprule
\includegraphics[width=0.95\linewidth,bb=5 5 1090 580]{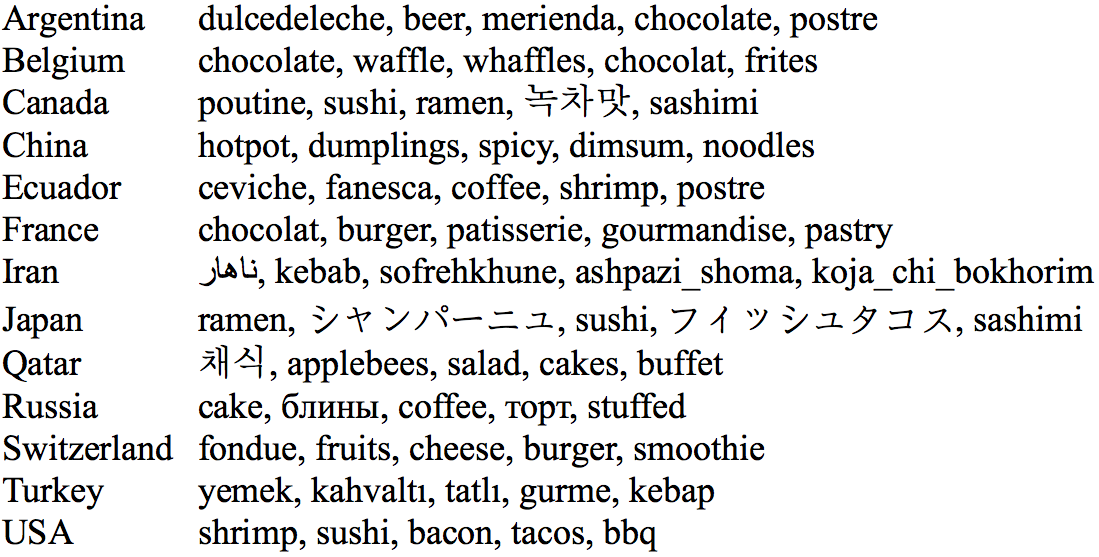}\\
\bottomrule
\end{tabular}
\end{center}
\end{table}

\section{National Dietary Behavior}
\label{sec:demogs}

Next, we expand our analysis to themes possibly signifying the dietary health and habits of the populations around the world. Emotions related to food intake have long been hypothesized to be associated with healthy weight maintenance~\cite{canetti2002food}, and social psychologists have studied food choice, weight control, and self-presentation~\cite{oconnor2004perceived}. Using the textual context of Instagram posts, we attempt to quantify emotional and social aspects of the favorite cuisines around the world.

\subsection{Hashtag Categories}

We begin by extracting the top hashtags of the posts associated with each of the 72 selected countries, and process them similarly to the previous section. First we get the top 1000 hashtags for each country, these are then converted to a probability distribution (by normalizing the term frequencies by their sum within the list). A similar vector is created for all countries by summing up all of the probability vectors and again normalizing by the sum to get a distribution of top 1000 hashtags. Then we normalize the top hashtag probabilities by the vector of most popular terms, to get the adjusted scores which would signify how \emph{prominent} the terms are for that particular country. To do this, we subtract the probability of a term appearing in ``all'' vector from its probability in the country vector. Note that here some scores may be 0, and some may end up being negative. We then sort the terms by this score
\footnote{the term distributions are available at \url{http://scdev5.qcri.org/sabbar/tags/foodporn.html}}.
The top term is invariably either the name of the country or its capital (or most famous, in case of New York or Barcelona, city). 


Now, we convert this qualitative data to quantitative. 
We use CrowdFlower to label the hashtags into categories dealing with social, health, and emotional aspects of dietary experience. We take the top 50 most distinguishing hashtags of each country, and exclude numbers and words of 1 or 2 characters. For each hashtag, links to Google Translate and Google Search were again provided, with instructions to select the last resort option of ``Don't understand'' only after attempting to translate the term. The following options were provided (as determined via content coding by the authors):


\begin{itemize}
\itemsep0em
\item Sentiment (emotions, opinions, etc)
\item Healthy (food, lifestyle, veggies, etc)
\item Unhealthy (food, lifestyle, desserts, etc)
\item Social (other people, events, holidays, etc)
\item Location (place, city, country, etc)
\item Food / drink (particular foods/drinks)
\item Time / date / time of the day / mealtime
\item None of the above
\item Don't understand
\end{itemize}

To ensure high quality of responses, we requested 5 different labels per hashtag. Twenty-three gold-standard questions were used to discard careless labelers and bots. Annotator agreement, as measured in the number of exact label overlaps, is 78.8\%, which is satisfactory for our task, considering there are 9 choices, and such that several can be selected at the same time.

The resulting 2060 labeled hash tags (available at \footnote{\url{https://tinyurl.com/foodporn-hashtags}}) have the distribution of labels shown in Table \ref{tbl:hashtagcategories}, along with some examples. Most tags are in English, but the set also includes many other languages and alphabets, including Korean, Cyrillic, and Chinese.

\begin{table}
\begin{center}
\caption{Categories of top distinguishing hashtags.}
\label{tbl:hashtagcategories}
\begin{tabular}{l|r|l}
\toprule
Category & \# & Examples \\ \midrule
Sentiment & 115 & yum, nomnom, blessed \\
Healthy & 106 & lowcarb, gym, goodeats \\
Unhealthy & 30 & fatty, cheatmeal, cake \\
Social & 114 & life, igers, girls \\
Location & 810 & italia, home, london \\
Food/drink & 657 & paella, pic\_food, mojito \\ 
Time/date & 52 & monday, evening, dinner \\
None of the above & 176 & vsco, tao, sun \\ 
\bottomrule
\end{tabular}
\end{center}
\end{table}

\subsection{National \& Regional Behavior}

\begin{figure}[t]
\centering
\includegraphics[width=\linewidth,bb=5 5 500 350]{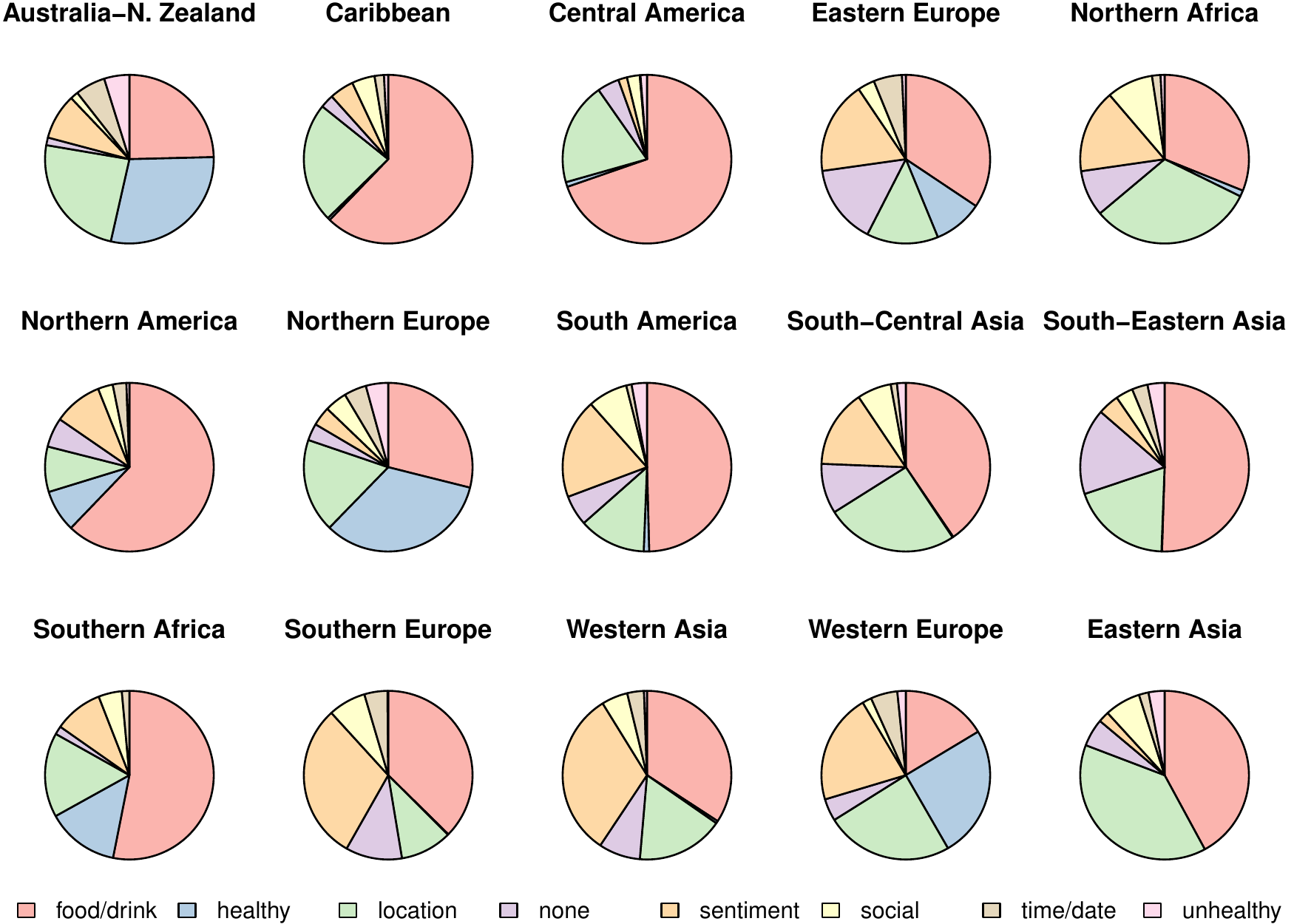}
\caption{Sub-regional hashtag category statistics}
\vspace{-0.3cm}
\label{fig:hashsubregion}
\end{figure}

Now that the hashtags have been categorized, the interests expressed through them in countries and their larger regions can now be quantified. For example, Figure \ref{fig:hashsubregion} shows the shares of each category in the top 50 hashtags of each part of the world. 

The most health-conscious regions are Northern and Western Europe, as well as Australia and New Zealand. Sentiment is most associated with \#foodporn in Southern Europe (Greece, Italy, Spain, etc.) and West Asia (Turkey and Middle East), and social events mentioned in Northern Africa (Egypt, Morocco), South America (Argentina, Brazil, etc.), South-Central Asia (India, Malaysia, etc.), and Southern Europe.

The highest rate of unhealthy tags came from Brazil, Argentina, and France. Examples of such popular tags are \#gordice (in Brazil), which derives from ``gordo'' (``fat''), and \#gourmandise (in France), which translates as ``gluttony''. These three countries have 5 healthy tags in their top 50 between them. 

The most healthy-conscious country is the Netherlands, with 25 out of the top 50 hashtags dealing with fitness and health, including \#fitgirl, \#fitspo, \#eatclean, etc. Ireland is a close second, with more bodybuilding-related tags, such as \#proteinpancakes, \#girlswholift, and \#fitness. 

As mentioned earlier, some countries display a higher population of tourists, and we find a sign of this trend in the social tags used in the two countries found to have the most of these: in El Salvador with \#family, \#friends, and \#sundayfunday and Cambodia  with \#wanderlust, \#instatravel, and \#holiday, among others. Similarly, United Arab Emirates is famous for Dubai and Abu Dhabi, destinations mentioned in no fewer than 17 out of 50 tags, but due to a high number of expatriate population, these may only partially indicate transient tourism.

\begin{figure}[t]
\centering
\includegraphics[width=0.8\linewidth,bb=5 5 500 350,trim={0cm 0cm 0cm 0.5cm},clip]{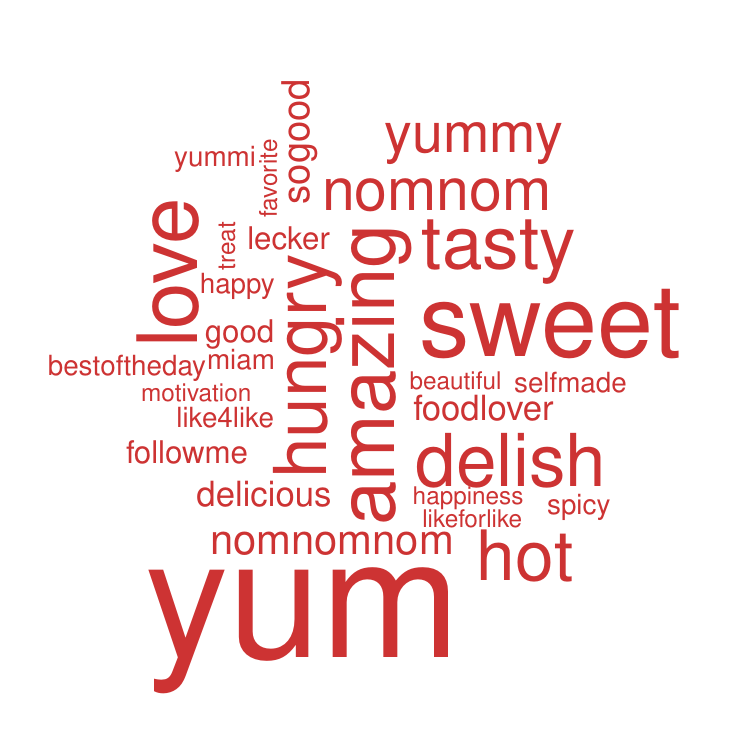} 
\vspace{-0.4cm}
\caption{Top emotions expressed in the context of \#foodporn hashtag.}
\label{fig:topemotions}
\end{figure}

Finally, we turn to the concern some social scientists have expressed over the use of ``porn'' in context of food consumption~\cite{mcbride2010food}. Although we find some unhealthy associations (which are sometimes highly localized), the emotions expressed around \#foodporn are overwhelmingly positive (see Figure \ref{fig:topemotions}). Generic emotions expressed include \#love, \#sogood, \#happiness, \#happy, \#good, etc. Among the top 30 emotions we also see \#motivation and \#selfmade, indicating association with healthy lifestyle.

\subsection{Demographic Correlation}

Whereas these hashtag categories provide a structured qualitative look at the associations with \#foodporn around the world, we ask whether there are quantitative relationships between the attitudes we glimpse and economic and social characteristics of each nation.
 
We use World Development Indicators (WDI)\footnote{\url{http://wdi.worldbank.org/tables}} data to enrich our understanding of these countries. The statistics we gathered are as follows:

\begin{itemize}
\itemsep0.3em
{\footnotesize
\item \emph{gdppc} - Gross Domestic Product per Capita
\item \emph{gini} - the Gini Index measuring income inequality
\item \emph{unemployment} - unemployment (\% of total labor force)
\item \emph{mobile} - mobile phone subscribers (per 1,000 people)
\item \emph{urban} - urban population (\% of total)
\item \emph{tourists} - international tourism, number of arrivals
\item \emph{diabetes} - diabetes prevalence (\% of population ages 20 to 79)
\item \emph{obesity} - from World Health Organization\footnote{\url{http://apps.who.int/gho/data/view.main.2450A}}
}
\end{itemize}

\begin{figure}[t]
\centering
\includegraphics[width=1.15\linewidth,bb=5 5 550 320,trim={0cm 0cm 0cm 1.5cm},clip]{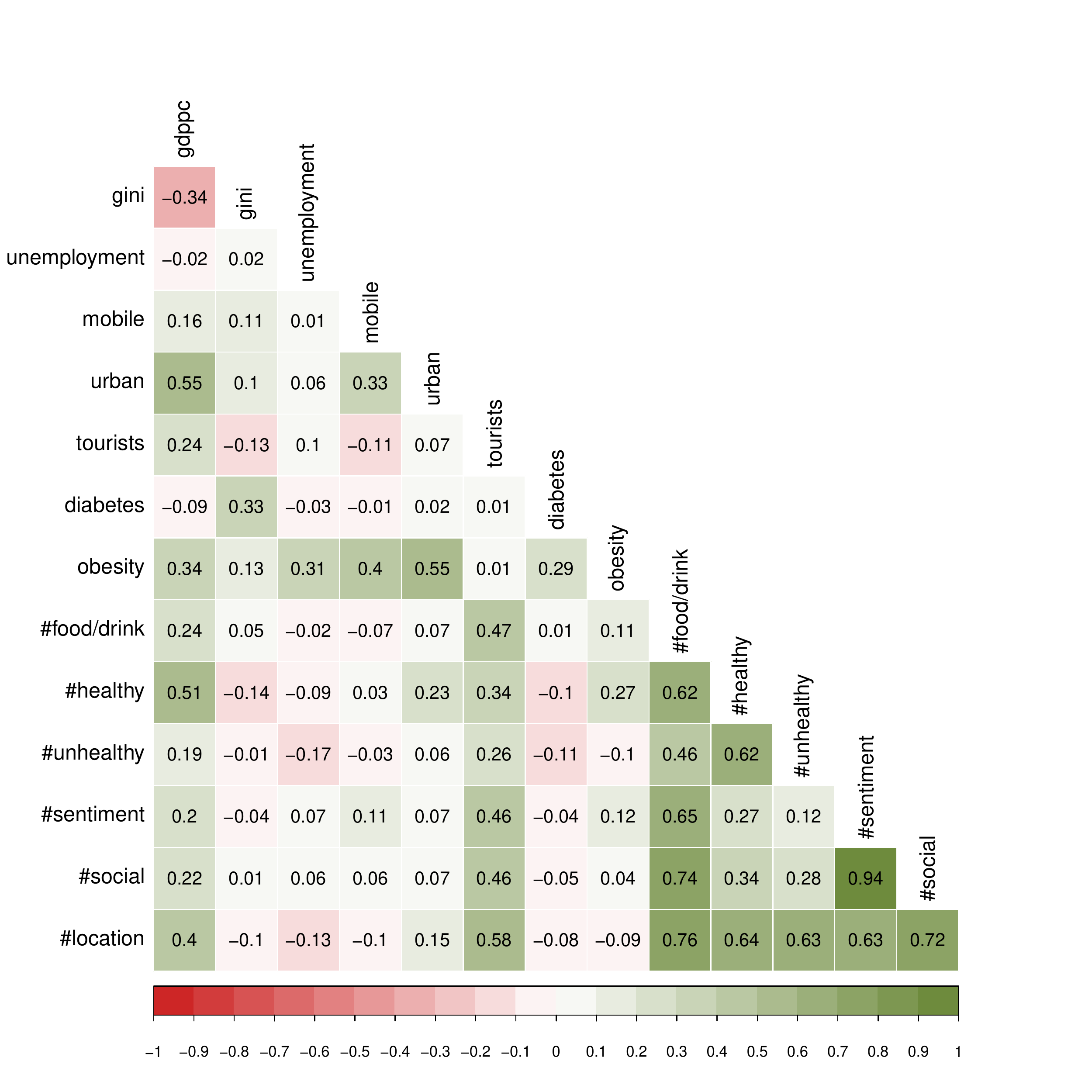} 
\caption{Correlation matrix of country demographics with hashtag categories (which are signified with \#).}
\label{fig:demogcorrel}
\end{figure}

Figure \ref{fig:demogcorrel} shows the correlation of the above social, economic, and health statistics with the hashtag use. The statistics begin the triangle from the top, and the hashtag statistics are aggregated per group and are signified with \# symbol. We see a few notable correlations -- especially those of GDP per capita (gdppc) with healthy (0.51) and location (0.40) hashtags. Healthy hashtags are also slightly negatively correlated with Gini index (at -0.14) such that the more unequal the incomes are (and the higher Gini index becomes), the fewer healthy tags are used. Interestingly, there is also a slight negative relationship between the use of unhealthy tags and unemployment rate (at -0.17). Tourism, we find, fairly strongly affects the use of all tags, but especially the mention of locations, social situations, and sentiment, as well as particular foods and drinks. 

Surprisingly, we find a positive correlation between obesity and mentions of healthy hashtags (at 0.27). This may be due to wealthy nations both having a greater problem with obesity, yet also enough disposable income and leisure time to concern with healthy lifestyle (perhaps in different populations). In order to discern the effects of different sub-populations, a more detailed profiling, either by automatic or standard survey techniques, is necessary. 

\section{Social Approval}
\label{sec:social}

\begin{figure}[t]
\centering
\includegraphics[width=0.9\linewidth,bb=5 5 330 220]{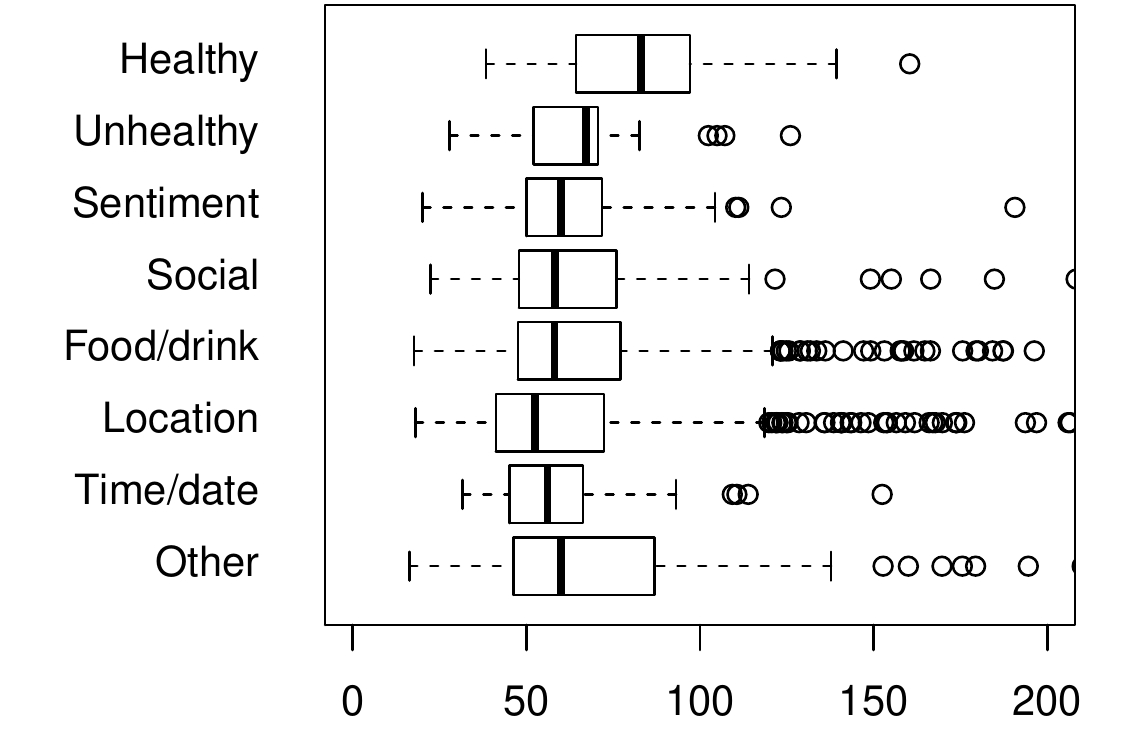}\hspace{0.7cm}
\caption{Distribution of average likes for posts containing hashtags from a certain group.}
\label{fig:classlikes}
\end{figure}

Finally, we turn to the social approval and interactions around \#foodporn. 
Here, we examine which kinds of behaviors are most promoted by the Instagram community. Figure~\ref{fig:classlikes} shows the distribution of average number of likes given to the posts containing a hashtag from one of the classes. To avoid extreme outliers from misspellings and singletons, we exclude all hashtags used by 4 or fewer unique users in our dataset. We then average over the number of likes each post containing a hashtag received. Healthy hashtags are by far the best liked, having an average of 87.6 likes (median 51), compared to 68.2 (median 35) of unhealthy ones (similarly, 4.7 comments for healthy and 3.7 for unhealthy). In fact, the top posted and liked hashtags in our dataset include \#eatclean, \#fresh and \#fitness.

\begin{figure}[t]
\centering
\includegraphics[width=0.9\linewidth,bb=5 5 330 220]{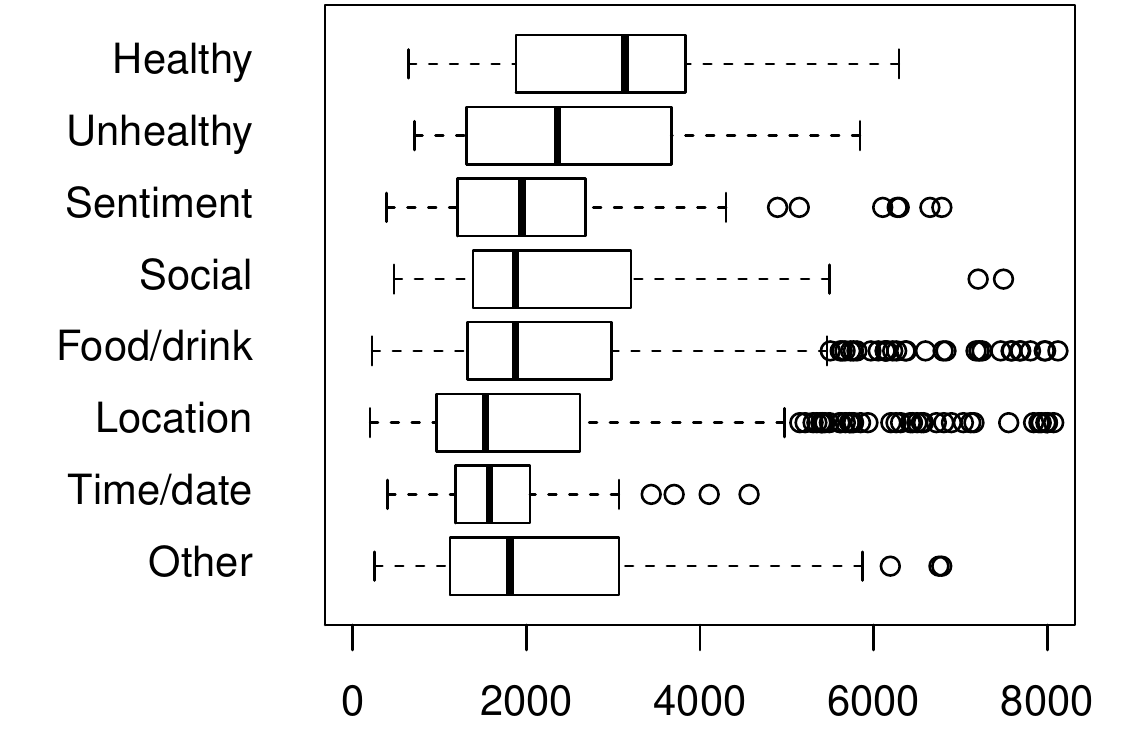}\hspace{0.7cm}
\caption{Distribution of average followers of users whose posts contain hashtags from a certain group.}
\vspace{-0.4cm}
\label{fig:classfollowers}
\end{figure}

However, it is the individual celebrities that receive the highest number of likes per post. At the top we find \#ztf, a hashtag referring to the fans of Zizan Razak, a Malaysian actor and comedian, with an average of 18,850 likes per post. The most commented tag is \#electrifynutrition, a supplement for weight training. These large communities adopt the \#foodporn hashtag and create their own context around it, redefining what their favorite food is (even when it is eaten by their favorite celebrity). 

Liking behavior is closely linked to the popularity of the posting user -- Pearson correlation between the likes of the posts and the number of followers of the posting user is 0.74. Indeed, the most popular Instagram users are those using the healthy hashtags, at an average of 3,426 followers, compared, for example, to 2,432 followers of users posting unhealthy hashtags (see Figure~\ref{fig:classfollowers}). It is unclear whether the prominence brings interest in health, or the health-related content to popularity, we detect a social approval of health-related content.


\section{Discussion}
\label{sec:discussion}


In this work we present a global view of the \#foodporn hashtag, as used in 72 countries around the world. We caution the reader from making interpolations about any particular country in this dataset. We show in Data Section that 15\% of posts are generated by users we dubbed as Travelers -- those posting in more than one country. As it was untenable to collect the posts of all the users and attempt to estimate their ``home'' country, we estimate that even more of these users may be traveling, and perchance not ``from'' the country to which they have been mapped. Additional bias comes from the English language of the query words (``food'' and ``porn'') -- albeit popular around the world, English use limits our access to the native languages. 

Regardless of query language, we do find a strong national cuisine popularity in many countries, including Japan, China, and Iran, with the accompanying alphabets and transliterations. The plurality of other nations, including Canada and USA, speak to the historical developments which resulted in a multi-cultural environment. Uniting most nations in our dataset is chocolate, so much so that a brand of chocolate spread -- Nutella -- has made it to the list of top 20 foods mentioned. This finding contrasts that obtained by the Oxfam survey~\cite{grow2011survey} of 17 countries around the world, who found pasta, Chinese, and pizza (in that order) to be the favorite. Although we find both pizza and pasta near the top of \#foodporn associations (in 3rd and 8th place, to be precise), our data presents a different definition of ``favorite'' foods. These differences may be due to the unique affordances social media presents to its users to form ``images'' of both themselves and their favorite foods. These may include the visual attractiveness of dishes, as well as the social context of online media sharing. 

Looking beyond chocolate and cake, we find that not all associations with \#foodporn are unhealthy. We find a positive correlation between a nation's obesity and the presence of healthy tags (see Figure \ref{fig:demogcorrel}), suggesting that in such communities there is a greater awareness of healthy lifestyle. Further, healthy hashtags' an association with GDPPC (GDP per capita) suggests the wealthier countries are developing the ``taste'' for healthy food, so much so that it is considered ``pornographic'' (a contradiction to its implied unhealthy connotations). Finally, the heightened social approval of healthy tags (Figure~\ref{fig:classlikes}) suggests that the community is already self-policing in promoting a healthier lifestyle. Stronger health- and fitness-oriented communities may play an important role, informing local policies for community health promotion. These results are also indicative of the ability to use social media for promoting health~\cite{neiger2012use}. In particular, this analysis informs the \emph{persuasive} technologies which incorporate experience formation, behavior tracking and reinforcement via big data and social media~\cite{fogg2002persuasive}.

Although \#foodporn may be considered playful or bombastic, its use may have real-world health implications for the individuals using it. Even though there is less stigma attached to consuming ``food porn'' than the non-food kinds, it may still impact individuals with eating disorders. Recently, Signe Rousseau~\cite{rousseau2014food} suggested that the perception that ``consuming food porn may be safer than consuming real food'' may especially effect the ``sufferers of eating disorders who rely on images of food as a substitute (within limits) for eating''. Further, much like pornography contributes to an unrealistic view of sexuality, an obsessive fixation on \#foodporn potentially promotes an ``unhealthy'' relationship with food. Our findings suggest that there is a diversity in the context in which the tag is used (including a healthy one), but a finer-grained analysis is required to detect the impact of \#foodporn on the users with eating disorders. 

\section{Conclusion}
\label{sec:conclusion}

In this paper we introduce a social media dataset capturing the global use of the highly popular \#foodporn hashtag on Instagram. At a high rate of geo-location, it captures the worldwide dietary trends much more thoroughly than the popular Twitter platform. Our analysis of these 9.3M posts reveals that while \#foodporn is often associated with high-calorie and sugary foods such as cake and chocolate, it also often appears in a healthy context. The sentiment associated with \#foodporn indicates that it is used to motivate healthy living, especially in countries with high GDPPC. This re-definition of ``gastro-porn'' within the social media community is another illustration of the effect social media has on our views and conceptualization of our surroundings and ourselves. 

Forthcoming collaboration with nutrition experts in the creation of rich, culturally diverse food libraries will be instrumental to a successful understanding of public health using the vast amount of data in social media. Resources similar to United States Department of Agriculture's food database\footnote{\url{http://ndb.nal.usda.gov/ndb/search/list}} should be extended to foods in other world cuisines for a quantitative comparison of nutrient balance (and imbalance) around the world. In ongoing efforts, we are also exploring computer vision techniques for understanding the content of images, including their composition and caloric value, for health and wellbeing research. 

\footnotesize
\bibliography{foodporn}
\bibliographystyle{aaai}
\end{document}